\documentclass[sigconf]{acmart}

\AtBeginDocument{%
  }

\copyrightyear{2025} 
\acmYear{2025} 
\setcopyright{cc}
\setcctype{by}
\acmConference[MICRO '25]{58th IEEE/ACM International Symposium on Microarchitecture}{October 18--22, 2025}{Seoul, Republic of Korea}
\acmBooktitle{58th IEEE/ACM International Symposium on Microarchitecture (MICRO '25), October 18--22, 2025, Seoul, Republic of Korea}\acmDOI{10.1145/3725843.3756048}
\acmISBN{979-8-4007-1573-0/2025/10}

\usepackage{multirow}
\usepackage{enumitem}
\usepackage{booktabs}
\usepackage{subcaption}
\usepackage{xspace}
\usepackage[linesnumbered,ruled,noend]{algorithm2e}
\usepackage{fancybox}
\usepackage{siunitx}

\newcommand{\tb}[1]{\textbf{#1}} 
\newcommand{\iname}[1]{\texttt{\textcolor{blue}{#1}}} 
\newcommand{\code}[1]{\texttt{#1}}

\newcommand{\name}{\textsc{Distributed-HISQ}\xspace}
\newcommand{\sync}{\textsc{BISP}\xspace}

\newcommand{\router}{router\xspace}

\newcommand{\squishlist}{
   \begin{list}{$\bullet$}
    {
    \setlength{\itemsep}{0pt}      \setlength{\parsep}{0pt}
      \setlength{\topsep}{3pt}       \setlength{\partopsep}{0pt}
      \setlength{\listparindent}{-2pt}
      \setlength{\itemindent}{-5pt}
      \setlength{\leftmargin}{1em} \setlength{\labelwidth}{0em}
      \setlength{\labelsep}{0.5em} } }

\newcommand{\squishend}{
    \end{list}  }

\newcounter{outercounter}
\newcommand{\squishenum}{%
  \begin{list}{\arabic{outercounter}}{%
    \usecounter{outercounter}%
    \setlength{\itemsep}{0pt}%
    \setlength{\parsep}{0pt}%
    \setlength{\topsep}{3pt}%
    \setlength{\partopsep}{0pt}%
    \setlength{\listparindent}{-10pt}%
    \setlength{\itemindent}{-2pt}%
    \setlength{\leftmargin}{1em}%
    \setlength{\labelwidth}{0em}%
    \setlength{\labelsep}{0.5em}%
    \let\makelabel=\makeplainlabel%
  }%
}
\newcommand{\makeplainlabel}[1]{#1.\hfill}
\newcommand{\squishenumend}{\end{list}}

\newcounter{innercounter}
\newcommand{\squishenumII}{%
  \begin{list}{\alph{innercounter}}{%
    \usecounter{innercounter}%
    \setlength{\itemsep}{0pt}%
    \setlength{\parsep}{0pt}%
    \setlength{\topsep}{3pt}%
    \setlength{\partopsep}{0pt}%
    \setlength{\listparindent}{-2pt}%
    \setlength{\itemindent}{-10pt}%
    \setlength{\leftmargin}{1em}%
    \setlength{\labelwidth}{0em}%
    \setlength{\labelsep}{0.5em}%
    \let\makelabel=\makenestedlabel%
  }%
}
\newcommand{\makenestedlabel}[1]{(#1)\hfill}
\newcommand{\squishenumIIend}{\end{list}}

\newcommand{\cz}{$\mathrm{CZ}$\xspace}

\newcommand{\xgate}{$X$\xspace}

\begin{document}\sloppy
\title{\name{}: A Distributed Quantum Control Architecture}

\author{Yilun   Zhao\textsuperscript{2,3,\ref{equal_contrib}}\quad 
Kangding Zhao\textsuperscript{1,\ref{equal_contrib}}\quad Peng Zhou\textsuperscript{4}\quad Dingdong Liu\textsuperscript{1}\quad Tingyu Luo\textsuperscript{5}\quad Yuzhen Zheng\textsuperscript{1}\quad Peng Luo\textsuperscript{1}\quad Shun Hu\textsuperscript{1}\quad Jin Lin\textsuperscript{6,7}\quad Cheng Guo\textsuperscript{6}\quad Yinhe Han\textsuperscript{2}\quad Ying Wang\textsuperscript{2}\quad Mingtang Deng\textsuperscript{1}\quad Junjie Wu\textsuperscript{1}\quad X. Fu\textsuperscript{1,\ref{corresponding_author}}
}

\affiliation{
  \institution{\textsuperscript{1}College of Computer Science and Technology, National University of Defense Technology, Changsha, China}
  \city{}
  \country{}
  \institution{\textsuperscript{2}Research Center for Intelligent Computing Systems, State Key Lab of Processors, Institute of Computing Technology, Chinese Academy of Sciences, Beijing, China}
  \city{}
  \country{}
  \institution{\textsuperscript{3}University of Chinese Academy of Sciences, Beijing, China}
  \city{}
  \country{}
  \institution{\textsuperscript{4}China Greatwall Technology Group Co., Ltd., Shenzhen, China}
  \city{}
  \country{}
  \institution{\textsuperscript{5}East China Normal University, Shanghai, China}
  \city{}
  \country{}
  \institution{\textsuperscript{6}Hefei National Laboratory, Hefei, China}
  \city{}
  \country{}
  \institution{\textsuperscript{7}University of Science and Technology of China, Hefei, China}
  \city{}
  \country{}
}


\renewcommand{\shortauthors}{Zhao \& Zhao, \textit{et al}.}




\newlength{\isawidth} 
\setlength{\isawidth}{\dimexpr\linewidth/80\relax}
\newlength{\pktwidth} 
\setlength{\pktwidth}{\dimexpr\linewidth/100\relax}
\newlength{\cachewidth} 
\setlength{\cachewidth}{\dimexpr\linewidth/160\relax}

\newcolumntype{C}[1]{@{}>{\centering\arraybackslash}m{#1}@{}}

\definecolor{myyellow}{HTML}{FBE59E}  
\definecolor{mygreen}{HTML}{85B243}    
\definecolor{mygreenlight}{HTML}{CEE55D}    

\newcommand{\insightbox}[2][1]{%
  \vspace{10pt}
  \noindent\fcolorbox{black}{mygreenlight}{%
    \parbox{\dimexpr\linewidth-2\fboxsep-2\fboxrule\relax}{%
      \textbf{Insight \##1}%
    }%
  }\\[-\fboxrule]
  \noindent\fcolorbox{black}{myyellow}{%
    \parbox{\dimexpr\linewidth-2\fboxsep-2\fboxrule\relax}{%
      \vspace{1mm}%
      #2%
      \vspace{1mm}%
    }%
  }%
  \vspace{10pt}
}



\begin{abstract}
The design of a scalable Quantum Control Architecture (QCA) faces two primary challenges.
First, the continuous growth in qubit counts has rendered distributed QCA inevitable, yet the non-deterministic latencies inherent in feedback loops demand cycle-accurate synchronization across multiple controllers.
Existing synchronization strategies --- whether lock-step or demand-driven --- introduce significant performance penalties.
Second, 
existing quantum instruction set architectures are polarized, being either too abstract or too granular. 
This lack of a unifying design necessitates recurrent hardware customization for each new control requirement,  which limits the system's reconfigurability and impedes the path toward a scalable and unified digital microarchitecture.

Addressing these challenges, we propose \name{}, featuring:  
(i) HISQ, A universal instruction set that redefines quantum control with a hardware-agnostic design. By decoupling from quantum operation semantics, HISQ provides a unified language for control sequences, enabling a single microarchitecture to support various control methods and enhancing system reconfigurability.
(ii) \sync{}, a booking-based synchronization protocol that can potentially achieve zero-cycle synchronization overhead.
The feasibility and adaptability of \name{} are validated through its implementation on a commercial quantum control system targeting superconducting qubits. We performed a comprehensive evaluation using a customized quantum software stack.
Our results show that \sync{} effectively synchronizes multiple control boards, leading to a 22.8\% reduction in average program execution time and a $\sim5\times$ reduction in infidelity when compared to an existing lock-step synchronization scheme.

\begingroup\renewcommand\thefootnote{*}
\footnotetext{\label{equal_contrib}Equal contribution.}
\endgroup

\begingroup\renewcommand\thefootnote{$\dagger$}
\footnotetext{\label{corresponding_author}email: \url{xiangfu@quanta.org.cn}}
\endgroup

\end{abstract}




\maketitle



\setcounter{section}{0}
\section{Introduction}

Memory, computation, and control are the three fundamental components of a programmable computer.
While qubits function as both memory and computational units for quantum information processing, an external dedicated control system is an indispensable component of a solid-state quantum computer.
This system serves to bridge the quantum computational core with the classical world accessible to end users by transmitting, receiving, and processing classical electromagnetic signals.

Both realizing quantum advantage using noisy intermediate-scale quantum systems~\cite{kingBeyondclassicalComputationQuantum2025a,liuCertifiedRandomnessUsing2025a} and achieving Fault-Tolerant Quantum Computing (FTQC)~\cite{acharyaQuantumErrorCorrection2025} necessitate a larger number of qubits,
with thousands or even millions of qubits expected~\cite{IBMroadmap_2025,GoogleRoadmap}.
This fact challenges the design of Quantum Control Architectures (QCA) with  simultaneously supporting   \textbf{programmability}\footnote{Programmability refers to the capability of flexibly defining the sequence of quantum operations applied on target qubits.},
\textbf{feedback}\footnote{Feedback refers to using the measurement result of one or more qubits to determine the following operations applied on the same or other qubits.},
and \textbf{scalability}\footnote{Scalability refers to the property that the quantum control system can satisfy the control requirements of a larger quantum system by duplicating basic modules with almost linear resource cost or lower.}.

QCAs can be roughly classified into centralized architecture or distributed architecture.
\textit{Centralized quantum control architectures}~\cite{fuExperimentalMicroarchitectureSuperconducting2017,fuEQASMExecutableQuantum2019,zouEnhancingNearTermQuantum2020,xiangSimultaneousFeedbackFeedforward2020,zhangExploitingDifferentLevels2021,xuQubiCOpenSourceFPGABased2021} feature a single binary executable with possible waveform configurations as input, whose digital part is usually implemented in a single FPGA chip.
Two factors significantly constrain the scalability of these centralized architectures.
Internally, the instruction issue rate associated with a single instruction cache or memory cannot afford the broader quantum operation stream required by the increasing number of qubits.
Externally, the limited number of pins available on a single FPGA chip sets an upper bound on the amount of analog channels controlling qubits.
To support the ever growing number of qubits, QCA inevitably adopts a distributed form.

\textit{Distributed Quantum Control Architectures} (DQCA) can offer improved scalability by partitioning a quantum program into multiple binaries executed across several control units.
Hence, it draws a lot of research attention and inspires various off-the-shelf products.
For example, QubiC 1.0~\cite{xuQubiCOpenSourceFPGABased2021}  utilizes a single FPGA board with 4 analog inputs and 4 outputs, which can only control no more than 4 superconducting qubits.
To increase control capacity, QubiC 2.0~\cite{fruitwalaDistributedArchitectureFPGAbased2024,xuQubiC20Extensible2023} adopts multiple RFSoC FPGAs, resulting in a DQCA.
The same rationale can be observed in other academic studies~\cite{huQuantumErrorCorrection2019,linScalableCustomizableArbitrary2019,yangFPGAbasedElectronicSystem2022,guoLowlatencyReadoutElectronics2022,zhangClassicalArchitectureDigital2024}, as well as off-the-shelf products that adopt board-level and/or chassis-level distribution~\cite{OPX,QCCS,ryanHardwareDynamicQuantum2017}.

By duplicating boards or chassis, these architectures offer improved scalability with higher instruction issue rates and more channels.
Nevertheless, these architectures either fail to simultaneously support scalability, programmability, and feedback~\cite{linScalableCustomizableArbitrary2019,huQuantumErrorCorrection2019,guoLowlatencyReadoutElectronics2022,yangFPGAbasedElectronicSystem2022}, or they largely overlook two critical requirements, namely (1) the challenge of reconciling cycle-level synchronization across controllers with the concurrent execution of multiple (potentially heterogeneous) binaries~\cite{zettles262DesignConsiderations2022,fruitwalaDistributedArchitectureFPGAbased2024,zhangClassicalArchitectureDigital2024}, and (2) the engineering effort required to adapt these architectures to diverse control requirements.
Before future technologies based on cryogenic CMOS or single-flux quantum devices get mature~\cite{minQIsimArchitecting10+K2023,byunXQsimModelingCrosstechnology2022,kimFaultTolerantMillionQubitScale2024,jokarDigiQScalableDigital2022}, these problems remain open challenges.

\subsection{Synchronization Challenge}
\label{ssec:sync_challenge}

Synchronization is essential for implementing multi-qubit operations (typically two-qubit gates).
Such operations require target qubits simultaneously go through particular state evolution, which in turn necessitates the concurrent application of control signals.
In a DQCA, control signals for the same operation may originate from different controllers.
Hence, it is crucial for these controllers to commit corresponding instructions at precisely the same time.

This synchronization requirement in quantum system is by nature different from that in classical computer systems.
In classical systems, synchronization primarily serves to guarantee expected data dependency across threads or processes.
The critical factor is often the relative order of operations, and waiting a few clock cycles rarely invalidates the fundamental logic.
In stark contrast, a timing error of even a few nanoseconds can lead to the  failure of a quantum gate~\cite{fuExperimentalMicroarchitectureSuperconducting2017}.
This corruption can further give rise to meaningless results.
Therefore, synchronization in DQCA demands instructions to be committed at almost the same wall-clock time, and can be as tight as at tens of picoseconds level.
This is not merely a preference for performance or a safeguard for logical order; it is a fundamental physical necessity rooted in the quantum mechanical evolution of the qubits.
Fortunately, this unprecedented synchronization requirement can be reduced to \tb{cycle-level instruction commitment synchronization\footnote{In this paper, we also use an equivalent term ``cycle-level instruction synchronization'', ``commitment'' is omitted for simplicity.}} in engineering practice, because it is feasible to synchronize the phase of different clocks at high precision via a clock distribution network, based on, e.g., phase-locked loops.

While lock-step synchronization—where controllers are synchronized at every clock cycle—is a straightforward approach, it introduces significant inefficiencies. In this scheme, if a controller executes instructions without knowledge of the other controllers' state, non-deterministic feedback operations can make it difficult to commit collaborative multi-qubit instructions that require precise timing. Enforcing lock-step synchronization incurs substantial execution overhead.
FFor instance, the approach used by the IBM system~\cite{zettles262DesignConsiderations2022} distributes the entire program flow to all controllers, with operations for other controllers replaced by  \emph{wait/idle/delay} instructions.
This design forces every feedback operation to incur data transmission across all controllers, leading to unnecessary communication and latency that scales with the number of feedback operations. Furthermore, as analyzed in QuAPE~\cite{zhangExploitingDifferentLevels2021}, this solution makes it difficult to execute simultaneous feedback operations on different qubits, which can significantly harm execution fidelity.

Another solution is to allow each controller to execute instructions at its own pace and synchronize them only when required.
This solution significantly reduces the amount of generated instructions, thereby increasing instruction execution efficiency.
Moreover, it may increase the flexibility of programming by allowing each controller to have its own control flow~\cite{fruitwalaDistributedArchitectureFPGAbased2024}.
To realize such an on-demand synchronization scheme,
an existing solution inserts a \emph{sync} instruction \emph{immediately} before the instruction to be synchronized, as adopted by Qubic 2.0~\cite{fruitwalaDistributedArchitectureFPGAbased2024}.
In this scheme, an unavoidable latency is introduced for the sync signal bouncing back.
In the context of quantum computing's pursuit of ultimate hardware performance, it is also highly desirable to eradicate this latency.

To summarize, it remains an open challenge to design a QCA that has an efficient and scalable synchronization scheme.

\subsection{Adaptability Requirement}

In contemporary quantum systems, the physical realization of qubits, control methodologies, and specific implementations of pulse generation and acquisition exhibit significant diversity and are undergoing rapid evolution.
For example, superconducting qubits differ structurally: some utilize fixed-frequency transmons~\cite{ibm_fixed_coupling_wei2024native}, while others incorporate tunable couplers to adjust gate speeds and enhance performance~\cite{acharyaQuantumErrorCorrection2025}.

For reasons of time and money,
it is of practical meaning for the digital part of the quantum control microarchitecture to support various gate implementation strategies to control various quantum chips with various analog implementation.
However, existing Quantum Instruction Set Architecture (QISA) designs are either too high-level or too low-level.
Some QISAs directly adopt quantum operations as instructions, which makes their microarchitectures difficult to support other quantum chip structures or operation implementation strategies~\cite{fuEQASMExecutableQuantum2019,zhangExploitingDifferentLevels2021}.
Instructions of other QISAs are tightly bound to specific control electronics, so that when adapting these architectures to another control electronic system, it inevitably leads to a clumsy and redundant microarchitecture implementation~\cite{ryanHardwareDynamicQuantum2017,xuQubiCOpenSourceFPGABased2021,xuQubiC20Extensible2023,huQuantumErrorCorrection2019,xiangSimultaneousFeedbackFeedforward2020,fuExperimentalMicroarchitectureSuperconducting2017}.
Therefore, it remains an open challenge to design an adaptable QISA.

\subsection{Contributions}

In this paper, we introduce \name{}, an architectural solution to systematically address the design requirements of DQCA with the following key contributions:
\begin{enumerate}
    \item HISQ, a \tb{H}ardware \tb{I}nstruction \tb{S}et for \tb{Q}uantum computing.
    We identify a new abstraction layer for adaptable QISA design, which is not only implementable on existing hardware but also expressive enough for potential applications.
    \item \sync{}, a \tb{B}ooking-based precise \tb{I}nstruction \tb{S}ynchronization \tb{P}rotocol.
    By introducing a two-condition synchronization method and advancing the sync instruction when possible, \sync{} can be implemented with low hardware cost.
    This approach enables both neighbor-level and region-level synchronization with  potentially zero-cycle overhead.
    When zero cycle is not possible, the overhead is still no more than that of existing synchronization schemes.

    \item We implement \name{} in a customizable commercial quantum control system. The adaptability of HISQ is validated by using the same HISQ core to control both the arbitrary-waveform-generator board and data-acquisition board. \sync{} has also been validated by multiple synchronization experiments using two boards with different instruction streams.

    \item We have designed and implemented a full quantum control software stack, which can describe and compile quantum algorithms and experiments into HISQ instructions. The feasibility of \name{} has been verified by various experiments on superconducting qubits with this control software stack.

\end{enumerate}

This paper is organized as follows. After Section~\ref{sec:consider} details the challenges of DQCA with corresponding insights, Section~\ref{sec:single-node} - \ref{sec:multi-node} introduces the single-node architecture, the synchronization protocol, and the distributed network of \name, respectively.
Verification and evaluation are performed in Section~\ref{sec:eval}.
After some discussion in Section~\ref{sec:discussion}, Section~\ref{sec:conc} concludes.

\section{Challenges and Insights}
\label{sec:consider}

In this section, we concretize challenges in  synchronization among controllers and adaptability of a hardware instruction set, and elaborate on how the challenges drive the design of \name{}.

\subsection{Synchronization}
\label{ssec:dist-consider}

\begin{figure}
    \centering
    \includegraphics[width=\linewidth]{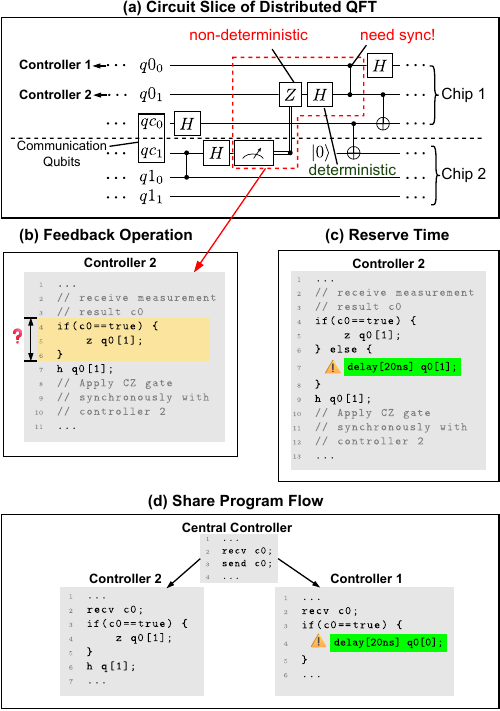}
    \caption{Motivational example of the synchronization challenge.
    (a) Example circuit slice derived from compiling QFT algorithm running on two quantum chips~\cite{diadamoDistributedQuantumComputing2021a,diskit}.
    (b) OpenQASM~\cite{crossOpenQASM3Broader2022} code snippets of the highlighted part in the circuit diagram.
    (c) Inserting a \emph{delay} instruction into the false branch.
    (d) Distribute the control flow to other controllers.}
    \label{fig:motiv-example-simple}
\end{figure}

\subsubsection{Origination of Synchronization Challenge}

The synchronization challenge of DQCA originates from the non-deterministic nature of \emph{dynamic quantum circuits}, which serve as critical components in many application scenarios.
For example, distributed quantum computing~\cite{magnardMicrowaveQuantumLink2020,diadamoDistributedQuantumComputing2021a} leverages quantum teleportation to connect multiple quantum chips to increase system scale,
long-range entanglements~\cite{baumerEfficientLongRangeEntanglement2024} can be created by dynamic circuits with lower circuit depth,
and logical T-gate in surface code~\cite{fowlerSurfaceCodesPractical2012} needs to be implemented via logical state teleportation.

Consider a system with multiple quantum chips connected via inter-chip quantum communication channels, with each qubit controlled by a distinct controller.
As illustrated in Figure~\ref{fig:motiv-example-simple}(a), a QFT algorithm running on this system relies on real-time feedback to implement cross-chip two-qubit gates~\cite{diskit,diadamoDistributedQuantumComputing2021a,wuAutoCommFrameworkEnabling2022} [highlighted in dashed box, with corresponding pseudo-code in Figure~\ref{fig:motiv-example-simple}(b)].
To achieve the minimal execution time, every gate should be executed as early as possible.
Since the execution of the $Z$ gate depends on a previous measurement result, the earliest time to execute the $H$ gate and following \cz gate becomes unpredictable at compile time.
However, the implementation of \cz requires synchronization between these two controllers,
which is challenging due to the unpredictable timing behaviors.

To solve this problem, we can either keep all the controllers synchronized at all time, or allow each controller to operate independently and synchronize with others only when necessary.

\subsubsection{Lock-Step Synchronization}
\label{sssec:drawback_all_the_time}

Based on whether there exist information exchange about control flow across controllers, two different methods can be used to achieve lock-step synchronization.
Unfortunately, both suffer from high execution overhead.

The first idea is to \tb{``reserve'' the time for the non-deterministic operation}, as seen in Figure~\ref{fig:motiv-example-simple}(c).
No matter the $Z$ operation is performed or not, the 20 ns time slot should be consumed.
The immediate drawback of this method is the introduction of ``dead time''
when the conditional operation is not performed.
Such unnecessary delay accumulates linearly as the number of feedback operations or the duration of conditional operation increases.
The longer execution time can degrade precious quantum fidelity.
More importantly, this solution cannot support repeat-until-success circuits with non-deterministic number of feedback loops~\cite{paetznickRepeatuntilsuccessNondeterministicDecomposition2014}.

\begin{figure}[hbt]
    \begin{subfigure}[b]{0.47\columnwidth}
        \centering
        \includegraphics[width=\textwidth]{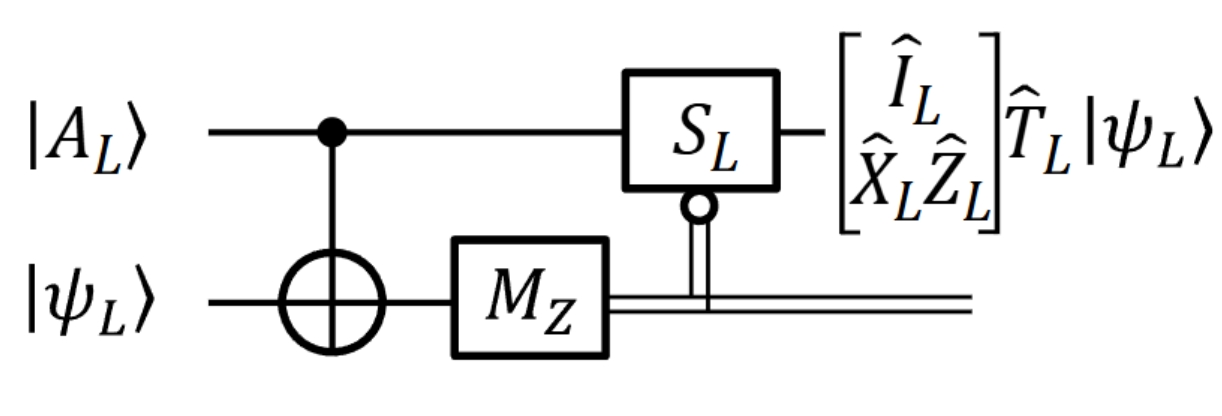}
        \caption{Logical $T$ gate.}
        \label{fig:first_image}
    \end{subfigure}
    ~~
    \begin{subfigure}[b]{0.47\columnwidth}
        \centering
        \includegraphics[width=\textwidth]{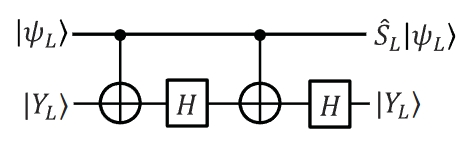}
        \caption{Logical $S$ gate.}
        \label{fig:second_image}
    \end{subfigure}

    \caption{(a) One implementation of logical $T$ gate relies on the conditional logical $S$ gate. (b) Logical $S$ gate is a sub-circuit with multiple logical operations that take a long execution time.}
    \label{fig:logical_t}
\end{figure}

The second idea is to \tb{``share'' the program flow information}, a strategy implemented in the IBM quantum control system~\cite{zettles262DesignConsiderations2022}
[Figure ~\ref{fig:motiv-example-simple}(c)].
When a measurement is performed, the outcome is broadcasted to all controllers so every controller proceed in lockstep down the same branch.
This solution is inherently limited by enforcing the same program flow on all controllers.
Firstly, this introduces unnecessary control flow and \emph{delay} instructions that may exaggerate the quantum operation issue rate problem limiting the scalability.
Secondly, it undermines programming flexibility as other controllers can only stay idle when some are performing feedback.
For example, it is difficult to realize simultaneous feedback~\cite{zhangExploitingDifferentLevels2021}. When the conditioned sub-circuit is long, temporally stacked feedbacks can accumulate much longer execution time than the parallelized version.
For example, some implementation scheme of the logical $T$ gate [Figure~\ref{fig:logical_t}(a)] relies on the conditional logical-$S$ gate [Figure~\ref{fig:logical_t}(b)], which in turn is a sub-circuit with multiple logical operations that can take a substantial execution time.~\cite{fowlerSurfaceCodesPractical2012}. If the feedback in logical $T$ gates can only be executed sequentially, the execution time of the entire program will grow significantly.
Ultimately, both issues can incur execution time overhead than expected that dampens program fidelity.

\subsubsection{Insights for Efficient Synchronization Protocol}

Given the substantial overhead of lock-step synchronization, an alternative is to  synchronize controllers in an as-needed manner.
This approach usually inserts a \emph{sync} instruction immediately before the instructions that requires cycle-level synchronization~\cite{fruitwalaDistributedArchitectureFPGAbased2024}.
Nonetheless, it will still introduce unavoidable latency for sync signals bouncing back.

In order to achieve the utmost synchronization efficiency, we draw inspiration from a common scenario in daily life: an individual organizing a meeting seeks to start it at the earliest possible time.
Each participant, upon determining their earliest available time, sends a message specifying this time to the organizer.
After collecting responses from all participants, the organizer identifies the latest of these times—representing the earliest feasible start for the meeting—and notifies all participants accordingly.
Provided the notification reaches participants before the designated start time, the meeting can proceed at the earliest time.

Since quantum operations usually have a fixed duration~\cite{krantzQuantumEngineersGuide2019}, a controller can potentially predict in advance the exact time at which it will reach a \emph{synchronization point}\footnote{The time point at which a controller finishes all operations before executing the operations that need be be synchronized.}, similar to a participant determining their earliest available time for a meeting.
Additionally, a common \router{} can serve as the meeting organizer.
It is possible for us to design a concise synchronization protocol based on the following insight:

\insightbox[1]{A controller can ``book'' a synchronization point  thanks to the deterministic nature of quantum operations.}

\subsubsection{Insights for Common Synchronization Scenarios}
\label{sssec:insight-comm-sync}

In the meeting example, the condition for the meeting to start at the earliest time is that participants receive the meeting time notification before its scheduled start.
For humans, the communication overhead is negligible.
However, a quantum controller must spend several cycles to communicate with other controllers.
Hence, to achieve the similar effect as meeting start, the communication latency must be suppressed as much as possible.

To meet this requirement, we identify two key characteristics of quantum applications and quantum devices which we can take advantage of:

\begin{enumerate}

\item Synchronization within a qubit region is a common scenario.
Quantum programs are often executed with multiple repetitions.
Before each repetition, it is usually necessary to perform a global synchronization among the involved controllers.
Considering that a quantum program is often mapped to a group of connected qubits (qubit region), the region-level synchronization becomes a common scenario.

\item Synchronization between controllers for neighboring qubits is a common scenario, since two-qubit gates are common cases and they can only be executed between physically adjacent qubits.
To minimize communication latency in these scenarios, the ideal situation is that these controllers are directly connected as neighbors.

\end{enumerate}

\insightbox[2]{Synchronization among controllers for a qubit region, and between controllers for neighboring qubits, are two common scenarios.}

These insights naturally give rise to a hybrid topology design with a simple routing mechanism to reduce communication latency, which will be detailed in Section~\ref{sec:multi-node}.

\subsection{Adaptability}
\label{ssec:isa-consider}

We can get a better understanding on the adaptability challenge of a QISA by digging into how a quantum operation should be implemented, or how the control electronics should behave.
Indeed, the hardware behavior of a quantum instruction can be affected by multiple factors.
\begin{itemize}[leftmargin=*]
    \item Different \textbf{qubit implementation technologies} require different control signals for the same quantum gate.
    For example, to implement a \cz gate, the corresponding control signals might be square pulses for superconducting qubits, while a set of modulated lasers for Rydberg atoms~\cite{sun2024BufferAtomMediated}.
    \item For the same qubit implementation technology, like superconducting qubits, different \textbf{quantum chip structures} require different control signals.
    Take the \cz gate as an example. Capacitor-coupled transmon qubits may require two square flux pulses and other possible auxiliary flux pulses to tune the frequencies of the target and adjacent qubits, respectively~\cite{dicarlo2009demonstration,versluis2017ScalableQuantumCircuit}. While tunable-coupler-coupled qubits can achieve the same effect using only three pulses~\cite{yan2018TunableCouplingScheme}.
    \item For the same quantum chip structure, different \textbf{operation implementation strategies} require the use of different control signals~\cite{zhang2025LearningForecastingOpen}.
    For the same coupler-coupled qubits as mentioned above, it is also possible to use all microwaves instead of square pulses to implement a \cz gate~\cite{li2022RealizationFastAllMicrowave}.
    \item Finally, even for the same control strategy of the same gate on the same qubits, different \textbf{implementations of the analog part} in the quantum control system can lead to different behaviors of the digital part.
    Taking the \xgate gate as an example, some systems may require the microarchitecture to trigger a pair of intermediate-frequency outputs from Digital-to-Analog Converters (DAC), which are later IQ-modulated with a radio-frequency carrier wave. On the contrary, direct-microwave-synthesis-based systems require to set the frequency and phase of the numerically controlled oscillator (NCO) and then directly trigger radio-frequency output from the DAC with a given envelope~\cite{kalfusHighFidelityControlSuperconducting2020} .
\end{itemize}
Since quantum chip fabrication, gate implementation strategies, and quantum control systems often advance at asynchronous pace, it is of ample economic and engineering significance for the same digital part of the quantum control microarchitecture to support various gate implementation strategies to control various quantum chips with various analog implementation.

However, it is difficult to adapt existing QISAs to various quantum chip structures and gate implementation strategies.
Some QISAs~\cite{fuEQASMExecutableQuantum2019,zhangExploitingDifferentLevels2021} directly adopt quantum operations as instructions, which are relatively too high-level for hardware implementation.
As a result, the corresponding microarchitecture can support the execution of these instructions targeting no more than one quantum chip structure without breaking the instruction definition.
For example, eQASM allows using two-qubit gates. However, its instantiated microarchitecture cannot support other quantum chips than the targeted seven-qubit chip.
In contrast, some other QISAs tightly bind the instruction semantics to the output channel behaviors~\cite{ryanHardwareDynamicQuantum2017,xuQubiCOpenSourceFPGABased2021,xuQubiC20Extensible2023,huQuantumErrorCorrection2019,xiangSimultaneousFeedbackFeedforward2020}.
The microarchitecture of these systems can hardly be adapted to other kinds of output channels or hardware behavior without sacrificing existing or adding new instructions or bit-fields with corresponding microarchitecture implementation. This fact usually leads to a clumsy microarchitecture implementation.
For example, many bit fields in QubiC~\cite{xuQubiC20Extensible2023} will be sacrificed if the definitions of instructions are altered to generate a marker signal.

Based on the above analysis, it is easy to comprehend that the key for designing an adaptable QISA is to figure out an abstraction layer, which is \textbf{not only implementable} on existing hardware, \textbf{but also expressive enough} for potential applications.
As quantum controllers are responsible for sending required control signals to expected qubits with correct timing,
we can extract the following insight for the instruction set design by decoupling quantum instructions from the operation semantics (top) as well as the concrete electronics behavior (bottom):

\insightbox[3]{At the instruction level, quantum control can be abstracted into: sending particular codewords, to particular ports, at particular time-points.}

By decoupling quantum semantics from hardware instructions, it significantly simplifies hardware implementation and allows this instruction to hook various underlying electronics implementations.
On the other hand, the quantum semantics can be taken care of another higher-level software instruction set, which can focus on the expressiveness of quantum computation and portability across various hardwares.

\begin{figure*}[ht]
\centering
\includegraphics[width=0.9\textwidth]{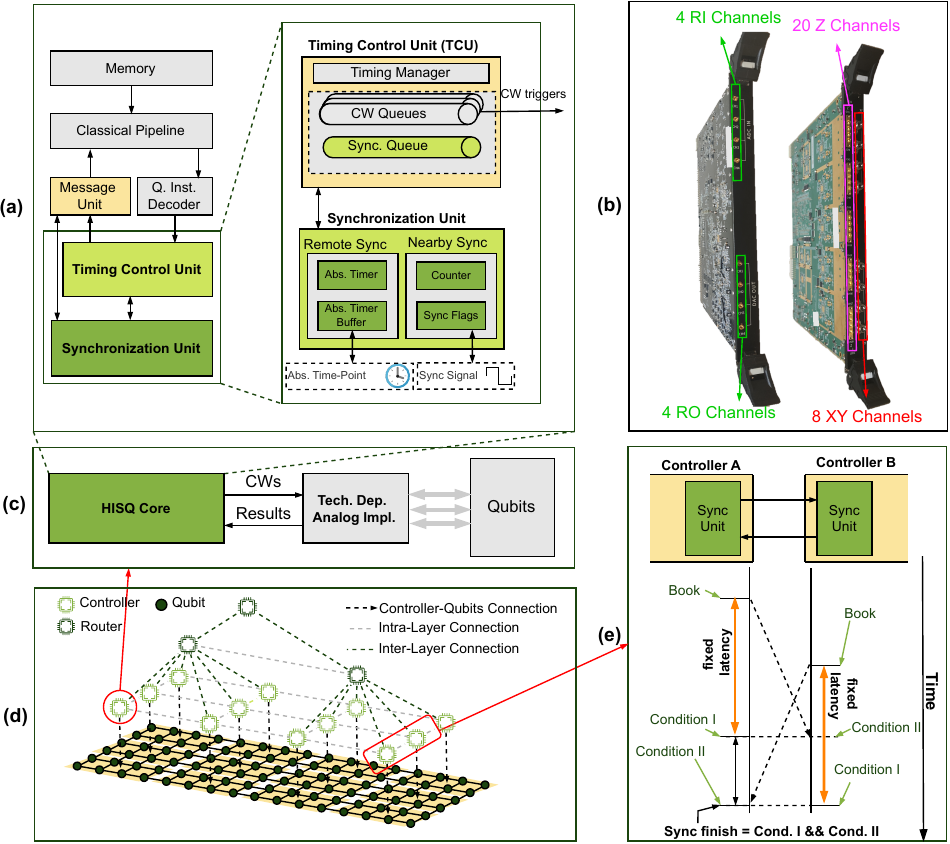}
\caption{Overview of \name{}. (a) Microarchitecture; (b) AWG board and readout board based on HISQ; (c) Single-node architecture; (d) Multi-node architecture;(e) Synchronization scheme.}
\label{fig:overview}
\end{figure*}

\section{Design Overview}
\label{sec:single-node}

Being a DQCA, \name{} comprises multiple control nodes coordinated by \router{}s [Figure~\ref{fig:overview}(d)].
A single node [Figure~\ref{fig:overview}(c)] integrates digital and analog components.
Its digital part, the HISQ core, abstracts quantum operations as ``sending particular codewords to particular ports at particular time-points''.
This abstraction decouples the instruction set from quantum semantics, allowing adaptation to diverse analog implementations.
To achieve efficient synchronization, the microarchitecture [Figure~\ref{fig:overview}(a)] builds on top of queue-based event timing control mechanism.
This mechanism can hide classical pipeline non-determinism and issue quantum instructions at precise timing.
By incorporating the meeting appointment insight, we design a highly efficient booking-based synchronization protocol (\sync{}) among control nodes [Figure~\ref{fig:overview}(e)], which can allow multiple controllers to re-sync their instruction execution at cycle-level precision after non-deterministic program subroutines.
This protocol is implemented by the newly introduced synchronization unit with modified timing control unit [Figure~\ref{fig:overview}(a)].
To enable efficient synchronization and feedback between remote controllers, \name{} employs a hybrid topology [Figure~\ref{fig:overview}(d)], which will be detailed in Section~\ref{sec:multi-node}.
Figure~\ref{fig:overview}(b) shows two types of boards implemented based on HISQ, which form the individual nodes of \name{}.
The detailed hardware implementation and configuration are described in Section~\ref{sec:eval}.

\subsection{Instruction Set Architecture}

\label{subsec:isa}

The core of designing a hardware-implementable instruction set is to provide the capability to describe operations of quantum controllers, including:
\begin{enumerate}
    \item Real-time classical register update and program flow control;
    \item Triggering particular actions at specific locations, including quantum operations and other behaviors like quantum error decoding;
    \item Timing control of controllers, including synchronization;
    \item Classical communication across controllers to support, e.g., feedback control.
\end{enumerate}
Hence, HISQ is designed to support the above four parts.

\subsubsection{Real-Time Classical Computation}

To reduce the burden in hardware and software implementation, HISQ is designed to be an extension to the RISC-V 32I instruction set.
To avoid disrupting timing behavior, we currently disable instructions/functionalities related to interrupts and memory fence.

\subsubsection{Triggering Operations}
\label{sssec:trigger_op}

As detailed in Section~\ref{ssec:isa-consider}, triggering operations can be abstracted as ``sending particular codewords, to particular ports, at particular time-points''.
To support flexibly specifying time-points, HISQ employs the timing control mechanism as proposed by QuMA~\cite{fuEQASMExecutableQuantum2019,fuExperimentalMicroarchitectureSuperconducting2017}.
Hence, both the immediate and register version of \iname{wait} instruction are included.

A set of codeword (\iname{cw}) instructions are used to describe ``sending particular codewords, to particular ports''.
The syntax of these instructions is:

\begin{center}
\fbox{\iname{cw.x.x}\code{ <port>, <codeword>}}
\end{center}

\noindent
Where, \iname{x} could be either \iname{i} or \iname{r}, indicating \code{<port>} and \code{<codeword>} specified by an immediate or a general-purpose register (GPR). For example, \iname{cw.i.r 3, r3} means sending the codeword specified by the GPR \code{r3} to the port No. 3.
The meaning of a codeword depends on the compiler and hardware configurations, thus it can vary in concrete implementations.
For example, a codeword can correspond to triggering a Gaussian pulse, setting the frequency of Numerically Controlled Oscillator (NCO), or any hardware action(s) that can be abstracted into a digital number.
A port may direct to the channels for I/Q, flux, readin/readout, or any other components that might need be controlled.
This port-based abstraction makes HISQ independent of the concrete implementations of operation/qubits while preserving sufficient expressiveness for controlling the hardware.

\subsubsection{Synchronization}
\label{sssec:sync_isa}

To support as-needed synchronization, HISQ includes a synchronization instruction with following syntax:

\begin{center}
    \fbox{\iname{sync} \code{ <tgt>}}
\end{center}

The \iname{sync} instruction can only function in combination with other \iname{sync} instructions, with each \iname{sync} instruction running on one controller.
The effect of a group of \iname{sync} instructions is to synchronize the clocks of controllers involved.

The field \code{<tgt>} is an immediate value, which can designates the address of (i) a controller or (ii) a router.
In the former case, \code{<tgt>} must refer to a nearest-neighbor controller, and there must be another \iname{sync} instruction running on the neighbor controller with its \code{<tgt>} field referring to this controller.
The result of both instructions is to
synchronize this controller and this neighbor controller.
In the latter case, \code{<tgt>} must refer to a ancestor router of this controller, and the effect is to synchronize with a subset of controllers managed by the same ancestor router.
For the sake of brevity, we will describe the synchronization mechanism in detail in Section \ref{sec:sync-protocol}.

\subsubsection{Classical Communication}

Classical information like measurement results should be transmitted across controllers to support real-time feedback and quantum error syndrome decoding.
To this end, HISQ includes \iname{send/recv} instructions, which are executed by a Message Unit (MsgU).

\subsection{Single-Node Microarchitecture}

\label{ssec:microarch}

The microarchitecture supporting HISQ is designed based on the QuMA microarchitecture~\cite{fuExperimentalMicroarchitectureSuperconducting2017,fuEQASMExecutableQuantum2019}. It differs from QuMA by slight modification [Figure~\ref{fig:overview}(a)] in the timing control unit (TCU) and the introduction of synchronization unit (SyncU) and the message unit (MsgU).

SyncU is used to support the \iname{sync} instruction and detailed in Section~\ref{sec:sync-protocol}.
Since sending and receiving messages across chips is well studied in previous network-related research, how MsgU supports \iname{send}/\iname{recv} is omitted in this paper for brevity.
Here, we mainly discuss the the basic working principle of TCU with required modification to support synchronization.

To achieve precise timing control, TCU employs the queue-based timing control mechanism~\cite{fuExperimentalMicroarchitectureSuperconducting2017}.
It contains a set of event queues with each corresponding to a port, where instructions can be  enqueued at non-precise timing while issued at designated precise timing.
The overall effect of TCU is to issue corresponding events at expected precise time-points.
We refer interested readers to Ref.~\cite{fuExperimentalMicroarchitectureSuperconducting2017,fuEQASMExecutableQuantum2019} for more details about queue-based timing control.

In the original queue-based timing control, there is no mechanism for TCU to wait for an external signal, and it can only support precise timing control in which all waiting durations are encoded into a fixed instruction sequence and calculated at runtime.
While synchronization is performed, one controller needs to wait for a non-deterministic duration depending on another controller, which cannot be supported by the original TCU.
Addressing this problem, TCU for HISQ is equipped with multiple ports receiving external triggers, that can be used to pause and resume the timer in TCU. In this way, TCU can allow events with non-deterministic timing while preserving precise timing control between non-deterministic-timing events.

\section{Synchronization Scheme}
\label{sec:sync-protocol}

\begin{figure}[b]
    \centering
    \includegraphics[width=\linewidth]{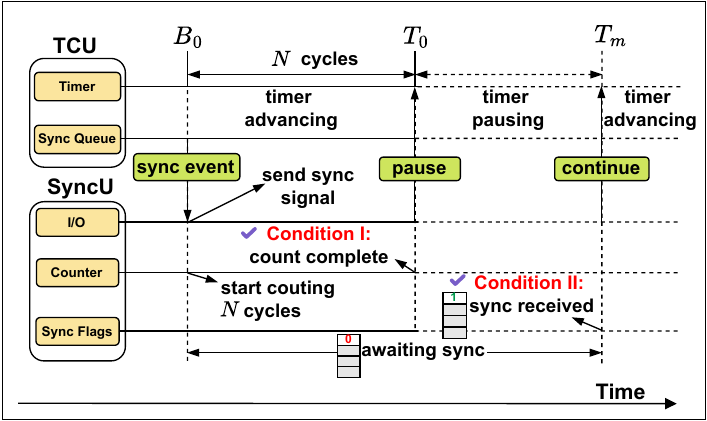}
    \caption{Single-node hardware behavior of \sync{}. The sync signal is exchanged with a neighbor controller. Upon receiving the signal, the corresponding sync flag (represented by stacked boxes for each neighbor) is set and cleared after being read.}
    \label{fig:sync-single-node}
\end{figure}

In a DQCA, feedback operations create dynamic timelines, making controller synchronization a key challenge (cf. Section~\ref{ssec:dist-consider}).
To achieve efficient synchronization, we introduce the \sync{} protocol.
We first demonstrate the simplicity of \sync{} through its single-node hardware behavior.
Then, we use examples to show the potential of \sync{} to achieve zero-cycle sync overhead in both nearby and remote synchronization scenarios.

\subsection{Node Actions to Achieve Synchronization}
\label{ssec:single-node-behavior}

The hardware implementation of the \sync{} protocol is very simple.
Taking the nearby synchronization between two adjacent controllers as an example, its single-node hardware behavior is as follows (Figure ~\ref{fig:sync-single-node}):

Synchronization begins when the TCU sends a sync event to the SyncU at the “booking” time $B_0$.
Upon receiving this event, the SyncU transmits a 1-bit signal to the target controller and starts an $N$-cycle countdown.
Here, $N$ is set to match the transmission delay between the hardware ports of the specific pair of neighboring controllers.
Since this delay is fixed and can be calibrated once the hardware connections are established, $N$ can be pre-configured in hardware for each connection.

Synchronization is achieved when both conditions are met:
\squishlist{}
    \item \tb{Condition I:} The $N$-cycle count completes.
    \item \tb{Condition II: } The sync signal from the target controller is received.
\squishend{}

\noindent
Note that when the $N$-cycle count completes, the SyncU may stall the TCU if the sync signal has not yet been received.
In this case, the TCU pauses executing quantum operations until the sync signal arrives;
otherwise, if the sync signal is received before the count completes, the TCU proceeds without interruption.

In our FPGA implementation, SyncU consumes only 13 LUTs.
\tb{Such a lightweight scheme not only ensures cycle-level instruction synchronization, but also can realize zero-cycle overhead in some scenarios.}

\begin{figure*}[h]
    \centering
    \includegraphics[width=\textwidth]{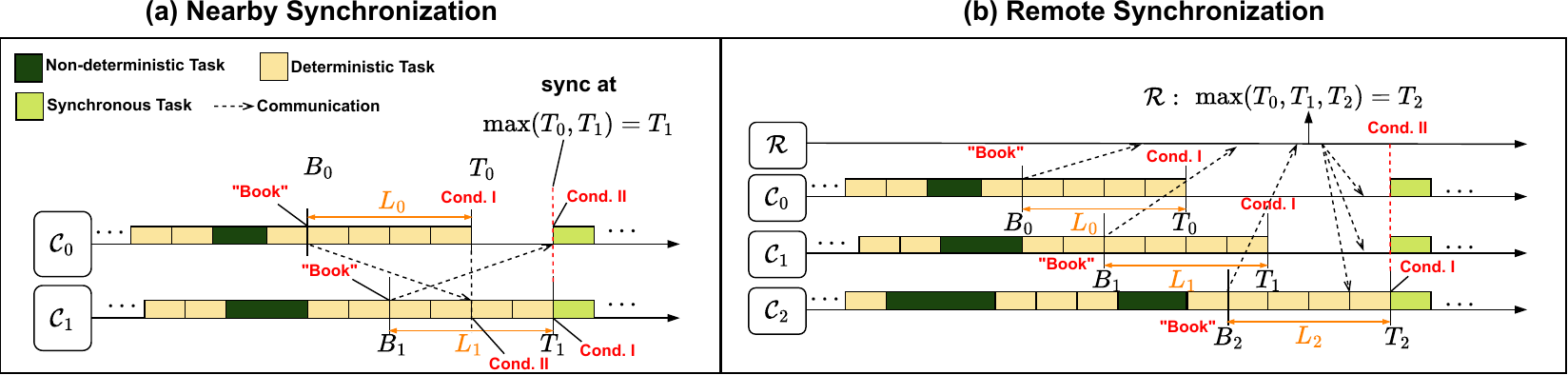}
    \caption{Example timing diagrams of nearby (a) and remote (b) synchronization using \sync{}.}
    \label{fig:sync-example}
\end{figure*}

\subsection{How Synchronization is Achieved}

\begin{figure}
    \centering
    \includegraphics[width=\linewidth]{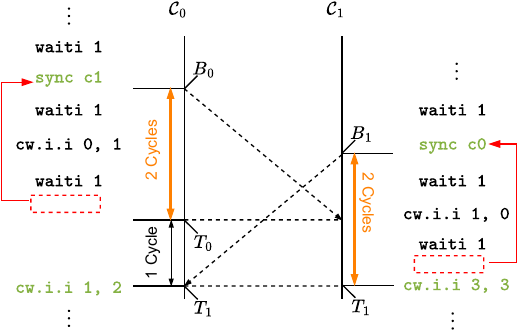}
    \caption{Example instructions for nearby synchronization.}
    \label{fig:inst-ahead}
\end{figure}

The effectiveness of \sync{} in nearby synchronization scenario can be observed from Figure~\ref{fig:sync-example}(a).

Firstly, we can observe that controllers $\mathcal{C}_0$ and $\mathcal{C}_1$ execute the synchronous task at the same time-point.
\squishlist{}
\item At the booking time-point $B_0$, $\mathcal{C}_0$ sends a sync signal to $\mathcal{C}_1$ and simultaneously starts counting for the signal transmission latency $L_0$.
At $T_0 = B_0 + L_0$, $\mathcal{C}_0$ has met \tb{Condition I}.
However, $\mathcal{C}_0$ does not receive the sync signal from $\mathcal{C}_1$ until $T_1$, and thus only meets \textbf{Condition II} at $T_1$. Therefore, it executes the synchronous task at $T_1$.

\item $\mathcal{C}_1$ sends its sync signal at $B_1$, which arrives at $T_1$, thereby satisfying \textbf{Condition I} at $T_1$.
Meanwhile, the sync signal arrives at $\mathcal{C}_1$ at $T_0$, satisfying \textbf{Condition II} at $T_0$.
Since $T_0 < T_1$, $\mathcal{C}_1$ satisfies both conditions and begins executing the synchronous task at $T_1$.

\squishend{}
Thus, both $\mathcal{C}_0$ and $\mathcal{C}_1$ execute the synchronous task at $T_1$.
In this diagram, $\mathcal{C}_0$ sends the sync signal before $\mathcal{C}_1$. If we swap $\mathcal{C}_0$ and $\mathcal{C}_1$ so that $\mathcal{C}_1$ sends the signal first, both controllers still begin executing the synchronous task at the same time.
This demonstrates that \sync{} ensures adjacent controllers start the synchronous task at the same time-point, thereby achieving cycle-level instruction synchronization.

Secondly, the \emph{synchronization overhead~\footnote{In this paper, synchronization overhead refers to the time interval between the last entity reaching the synchronization point and the first entity beginning the synchronous task.}} is potentially ``zero-cycle''.
For both $\mathcal{C}_0$ and $\mathcal{C}_1$, assume there are several \emph{deterministic tasks} (light-yellow blocks) between their last \emph{non-deterministic task~\footnote{A non-deterministic task refers to a task with unpredictable duration, e.g.,  a feedback operation. Similarly, a deterministic task has a fixed duration, e.g., a quantum gate.}} (dark green) and the synchronous task (light green).
The time-points that $\mathcal{C}_0$ and $\mathcal{C}_1$ finish these deterministic tasks are $T_0$ and $T_1$, respectively.
Hence, the earliest time that $\mathcal{C}_0$ and $\mathcal{C}_1$ can co-execute the synchronous task is given by $\max(T_0,T_1) = T_1$.
Consider both $\mathcal{C}_0$ and $\mathcal{C}_1$ execute the synchronous task at exactly $T_1$, we can consider that \sync{} achieves zero-cycle overhead in this case.

The core idea of \sync{} lies in the "booking" mechanism.
As long as there are deterministic tasks with sufficient duration to cover communication latency, we can book a synchronization point in advance.
\tb{This allows us to insert a \iname{sync} instruction ahead of the synchronization point (Figure~\ref{fig:inst-ahead}), rather than placing it immediately before the synchronization point as done in Qubic~\cite{fruitwalaDistributedArchitectureFPGAbased2024}.}
Consequently, \sync{} minimizes the synchronization overhead (cf. \hyperref[insight:meeting]{Insight 1}).
This approach can also be easily extended to more complex remote synchronization scenarios.

\subsection{Extending to Remote Synchronization}
\label{ssec:remote-sync}

Figure~\ref{fig:sync-example} exemplifies a remote synchronization scenario, where controllers $\mathcal{C}_0,\mathcal{C}_1,$ and $\mathcal{C}_2$ synchronize with each other through a  \router{} $R$.

These controllers, whether adjacent or not, can still achieve synchronization with zero-cycle overhead.
As in the nearby synchronization case, we assume each controller has deterministic tasks preceding the synchronous task.
Each controller sends its earliest possible start time ($T_0$, $T_1$, or $T_2$) to $R$ before completing its deterministic tasks.
Once $R$ receives all synchronization requests, it determines and broadcasts the earliest common start time, $T_m = \max(T_0, T_1, T_2)$, to all controllers.
Consequently, all controllers begin executing the synchronous task at $T_m$, achieving cycle-level synchronization at the earliest possible time with zero-cycle overhead.

After sending $T_0/T_1/T_2$, each controller $\mathcal{C}_i$ achieves synchronization when both of the following conditions are met:
\squishlist{}
\item \tb{Condition I}: The current absolute time reaches $T_i$.
\item \tb{Condition II}: The earliest sync point $T_m$ has been received.
\squishend{}

\noindent
At this point, the SyncU checks whether the current time [indicated by the Abs. Timer shown in Figure~\ref{fig:overview}(a)] has reached $T_m$ [stored in the Abs. Timer Buffer shown in Figure~\ref{fig:overview}(a)], pausing the TCU’s timer if necessary and resuming it precisely at $T_m$.

\subsection{Condition for Zero-Cycle Overhead}

\begin{figure}
    \centering
    \includegraphics[width=\linewidth]{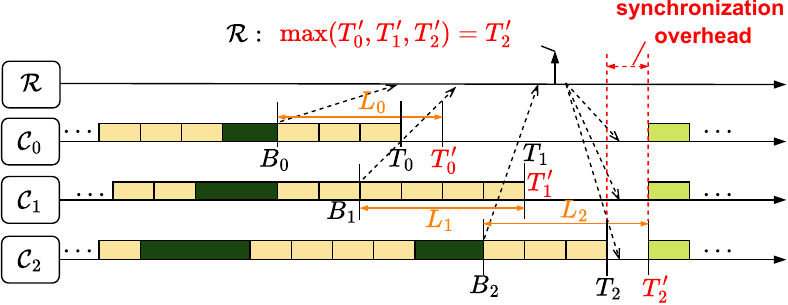}
    \caption{Example timing diagram of remote synchronization with non-zero overhead.}
    \label{fig:sync-overhead-example}
\end{figure}

In real-world scenarios, there may not always be a sufficient number of deterministic tasks available prior to a synchronization task to mask the communication latency.
In such scenarios, \sync{} may still introduce synchronization overhead.

In the example shown in Figure~\ref{fig:sync-overhead-example}, all controllers can accomplish their tasks before synchronization at $T_2$, which forms the theoretical earliest synchronization time.
However, the duration of deterministic tasks of $\mathcal{C}_2$ is $D_2 = T_2 - B_2 < L_2$, which is insufficient to hide the sync communication latency $L_2$ between $\mathcal{C}_2$ and $R$.
Thus all controllers involved cannot synchronize until $T_2^\prime$, resulting in a synchronization overhead of $L_2 - D_2$.

In this case, the actual earliest possible start time for $\mathcal{C}_0,\mathcal{C}_1,\mathcal{C}_2$ is $\max(T_0^\prime,T_1^\prime,T_2^\prime) = T_2^\prime$, which exceeds the theoretical earliest start time $T_2$.
As a result, achieving zero-cycle overhead is impossible.
More generally, we can conclude that zero-cycle overhead can be realized if and only if the actual earliest start time is the same with the theoretical earliest start time, formally expressed as
$\max(\{B_i + L_i\}) = \max(\{T_i\})$.

\section{Distributed Architecture}

\label{sec:multi-node}

As a distributed QCA, \name{} requires a network topology tailored for quantum applications to minimize communication overhead that could impair architectural scalability.
In this section, we first describe the topology design of \name{}, followed by an illustration of the \router{} design, emphasizing its routing mechanism.

\subsection{Design of Topology}
\label{subsec:topology}

We design a hybrid topology in \name{}, which consists of a tree-like inter-layer topology and a mesh-like intra-layer topology as shown in Figure~\ref{fig:overview}(d).
The controllers at the bottom layer are coordinated by higher-level \router{}s.

The tree-like inter-layer topology is adopted to minimize network edges while also reducing communication hops at region-level.
For a connected graph with $N$ nodes, the minimum number of edges is $N-1$, forming a tree since each controller has finite connections.
The network’s maximum communication latency depends on the topology graph’s diameter, which in a tree is $2 \times h$, where $h$ is the tree height.
Hence, a balanced tree with minimal height is adopted to reduce latency.

Additionally, \hyperref[insight:nearby]{Insight 3} indicates that the intra-layer topology should mirror the qubit device topology, naturally resulting in a mesh-like structure.

\subsection{Routing Mechanism}

Remote synchronization at region-level and nearby synchronization, are two common scenarios (Section~\ref{sssec:insight-comm-sync}).
While nearby synchronization involves only direct communication between neighboring controllers, region-level remote synchronization requires \router{}s with efficient routing mechanisms to reduce communication latency.

\begin{figure}[bt]
    \centering
    \includegraphics[width=0.6\linewidth]{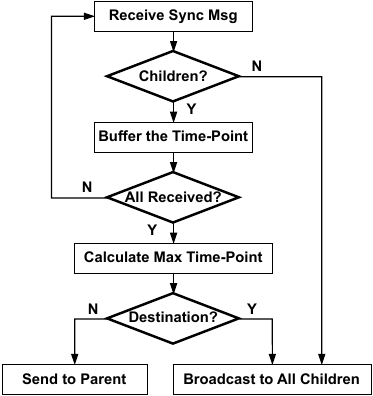}
    \caption{Router actions of region-level synchronization.}
    \label{fig:routing}
\end{figure}

The \router{} incorporates a simple routing mechanism that leverages the nature of tree topology (Figure~\ref{fig:routing}).

\begin{enumerate}
    \item Upon receiving a message, it buffers the message if it is from a child; otherwise, it broadcasts the message to all children.
    \item After receiving messages from all children, it computes the maximum time-point (Section~\ref{ssec:remote-sync}).
    \item If the message is destined for itself, it broadcasts the maximum time-point to all children; otherwise, it sends the time-point to its parent.
\end{enumerate}

\section{Evaluation}
\label{sec:eval}
\vspace{-2pt}

\subsection{Hardware Implementation}

\begin{figure}
    \centering
    \includegraphics[width=0.7\linewidth]{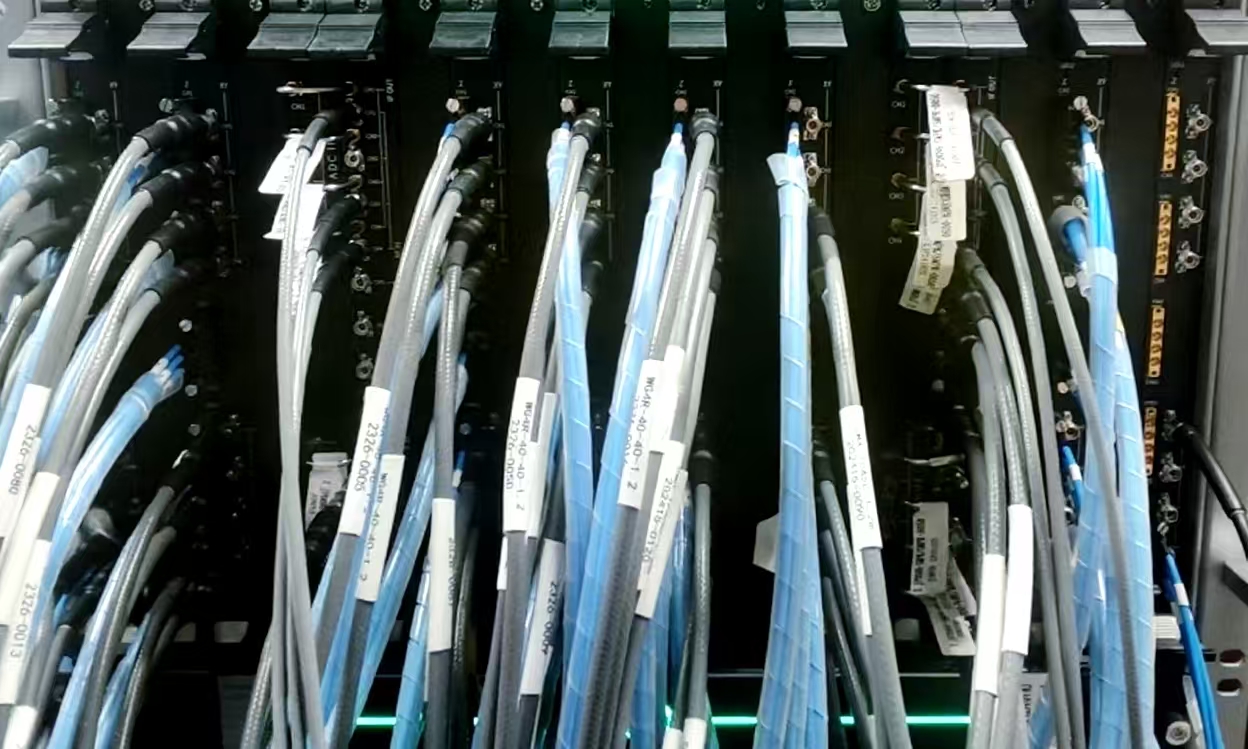}
    \caption{Hardware implementation of \name{}.}
    \label{fig:multi-boards}
\end{figure}

We implement \name in a commercial distributed quantum control system (named DQCtrl, Figure~\ref{fig:multi-boards}).
The control boards and readout boards, as shown Figure~\ref{fig:overview}(b), form the leaf nodes of \name{}.
The control board has eight XY channels for $x/y$ rotations and 20 Z channels for flux control.
The readout board comprises four pairs of input and output channels.
Each board is controlled by a single HISQ core.
All control boards and readout boards are connected using a back-plane, through which, the readout board can connect to each control board, each with two Low-Voltage-Differential-Signal (LVDS) channels and a dedicated channel for global trigger distribution.
Due to the limited connectivity, we are not able to realize region-level synchronization.

The same microarchitecture as shown in Figure~\ref{fig:overview}(a) is deployed on both the control and readout board, with the only difference between them being the number of codeword queues, which matches the amount of channels on each board.
The same HISQ instruction set is used to control all the XY channels and Z channels on the AWG boards, as well as measurement excitation and data acquisition on the readout boards.
Nevertheless, the same codeword can produce entirely different behaviors on different boards.
For example, the instruction \iname{cw.i.i}\code{ 1, 1} on the AWG applies an X gate, while on the readout board it triggers a measurement result discrimination.
Such kind of implementation is a simple verification of the adaptability of HISQ.


\begin{table}[bt]
\centering
\caption{Overview of FPGA resource consumption of HISQ on the control and readout board.}
\label{tab:fpga_resource}
\footnotesize 
\begin{tabular}{@{} p{2.5cm}  >{\centering\arraybackslash}p{1cm} >{\centering\arraybackslash}p{1.5cm} >{\centering\arraybackslash}p{1cm} @{}}
\toprule
Type & \#LUTs & \#Block RAM (32Kb~per~block) & \#FF \\
\midrule
\multirow{1}{2.5cm}{Control Board} 
  & 4,155 & 75 & 6,392 \\
\midrule
\multirow{1}{2.5cm}{Readout Board} 
  & 2,435 & 45 & 3,192 \\
\midrule
\multirow{1}{3cm}{Event Queue (38bit~x~1024)} 
  & 86 & 1.5 & 160 \\ 
\bottomrule
\end{tabular}
\end{table}
The HISQ implementation is characterized by its high efficiency in FPGA resource utilization, as detailed in Table~\ref{tab:fpga_resource}. A full 28-channel control board requires only 4155 LUTs, 6392 FFs, and \SI{2.46}{Mb} of Block RAM, while an 8-channel readout board uses 2435 LUTs, 3192 FFs, and \SI{1.47}{Mb} of Block RAM. For precise timing control, the TCU operates at \SI{250}{\MHz}, enabling a \SI{4}{\ns} resolution grid that improves upon the classic \SI{200}{\MHz} pipeline based on the PicoRV32I core~\cite{picorv32}.

In the context of modern FPGAs, which offer vast logic and memory resources, the hardware cost of HISQ is minimal. Its compact and efficient design ensures high scalability and straightforward portability to other systems without concerns about logic complexity becoming a bottleneck.

\subsection{Qubit-Level Verification}

\begin{figure}[bt]
    \centering
    \includegraphics[width=0.8\linewidth]{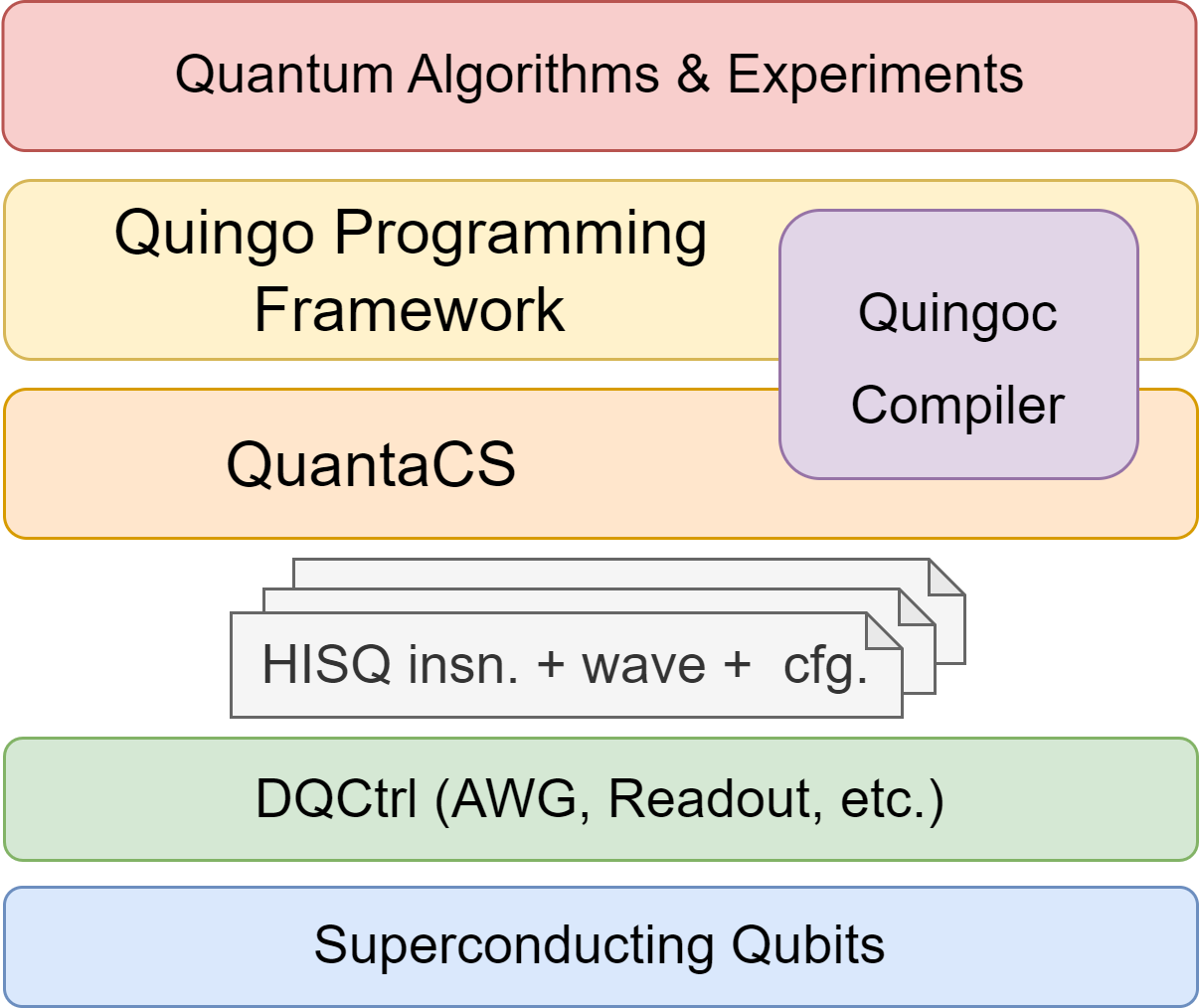}
    \caption{Quantum software stack for \name{}.}
    \label{fig:full_stack_verify}
\end{figure}

To validate \name{} on physical qubits, we developed a full quantum software stack (Fig.~\ref{fig:full_stack_verify}) comprising the Quingo programming framework~\cite{fuQuingoProgrammingFramework2021}, the MLIR-based Quingoc compiler, QuantaCS control software, and the SISQ instruction set. The Quingo framework integrates Python for data analysis with its native quantum language, which supports both high-level algorithms and low-level calibrations. Quingoc compiles Quingo programs into circuit-layer SISQ, which is then lowered to a hardware-agnostic, pulse-level representation. Finally, a new compiler backend for DQCtrl partitions the pulse-level program, generating HISQ binaries, waveform tables, and configuration files for the control hardware.

The target device is a superconducting quantum chip with 66 qubits and 110 couplers. 
Qubit operating frequencies range from \SIrange{3.953}{4.757}{\GHz}, with readout frequencies ranging from \SIrange{6.220}{6.560}{\GHz}. 
66 XY channels, 176 Z channels, and 11 readout channels with each capacitively coupling 6 qubits, are used to control and measure these qubits.

\begin{figure}[bt]
    \centering
    \includegraphics[width=\linewidth]{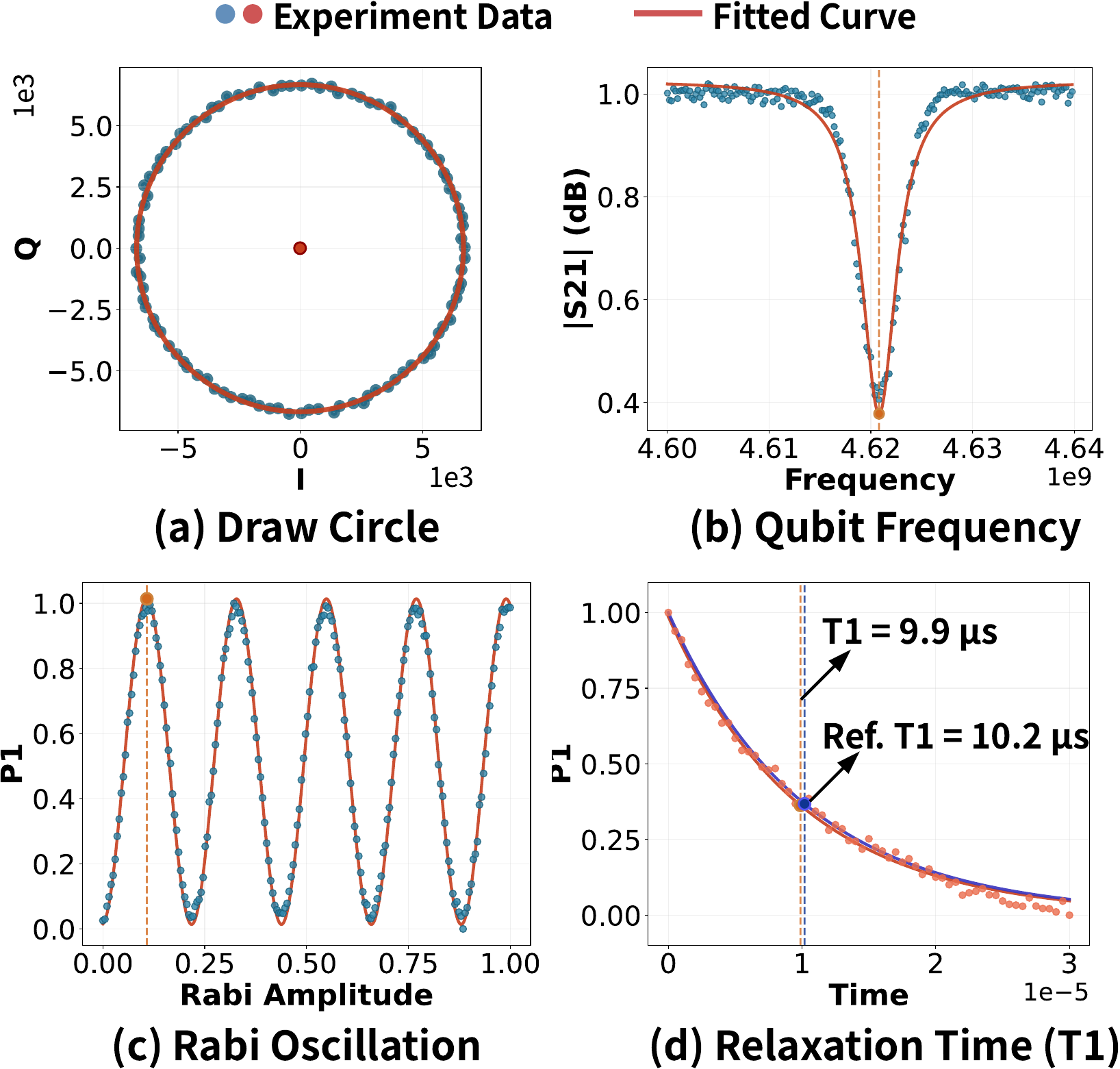}
    \caption{Four calibration experiments performed on a superconducting qubit, which show the capability of \name{} controlling signal (a) phase, (b) frequency, (c) amplitude, (d) timing and (a,c,d) pulse envelop.}
    \label{fig:qubit}
\end{figure}

We conducted a series of calibration experiments on multiple superconducting qubits to demonstrate the capabilities of \name{}. Figure~\ref{fig:qubit} presents four selected experiments, each designed to characterize a fundamental property of the control signals.
Figure~\ref{fig:qubit}(a) shows a self-verification experiment for the readout board. A measurement excitation pulse with a linearly increasing \textbf{phase} is emitted, and the response is collected, IQ-demodulated, and integrated. This process yielded a characteristic circular pattern in the IQ plane. The observed deviation from an ideal circle arises from small but non-negligible interference from adjacent qubits coupled to the same feedline.
Figure~\ref{fig:qubit}(b) depicts a spectroscopy experiment to determine the qubit's operating frequency. An $x$-rotation pulse was applied at varying \textbf{frequencies}, followed by a measurement of the qubit state, identifying the qubit resonance at \SI{4.62}{GHz}.
Figure~\ref{fig:qubit}(c) displays a Rabi oscillation experiment, performed to find the optimal pulse \textbf{amplitude} for implementing a high-fidelity $X$ gate.
Figure~\ref{fig:qubit}(d) presents a measurement of the qubit relaxation time ($T_1$), which characterizes the decay of the qubit’s excited state population over \textbf{time}. This measurement yielded a relaxation time of \SI{9.9}{\us}. 
For comparison, we also measured the working frequency and relaxation time of the same qubit using identical hardware but with an alternative and more mature firmware and software stack, obtaining values of \SI{4.64}{GHz} and \SI{10.2}{\us}, respectively. 
The minor discrepancies are well within the expected range attributable to the natural temporal fluctuations of the qubit's state.
Collectively, the fact that all experiments generated data conforming to their expected theoretical patterns demonstrates that \name{} can produce high-fidelity qubit control signals with precise, real-time manipulation of phase, frequency, amplitude, and timing.

\subsection{Electronics-Level Verification}

\begin{figure}[bt]
    \centering
    \includegraphics[width=\linewidth]{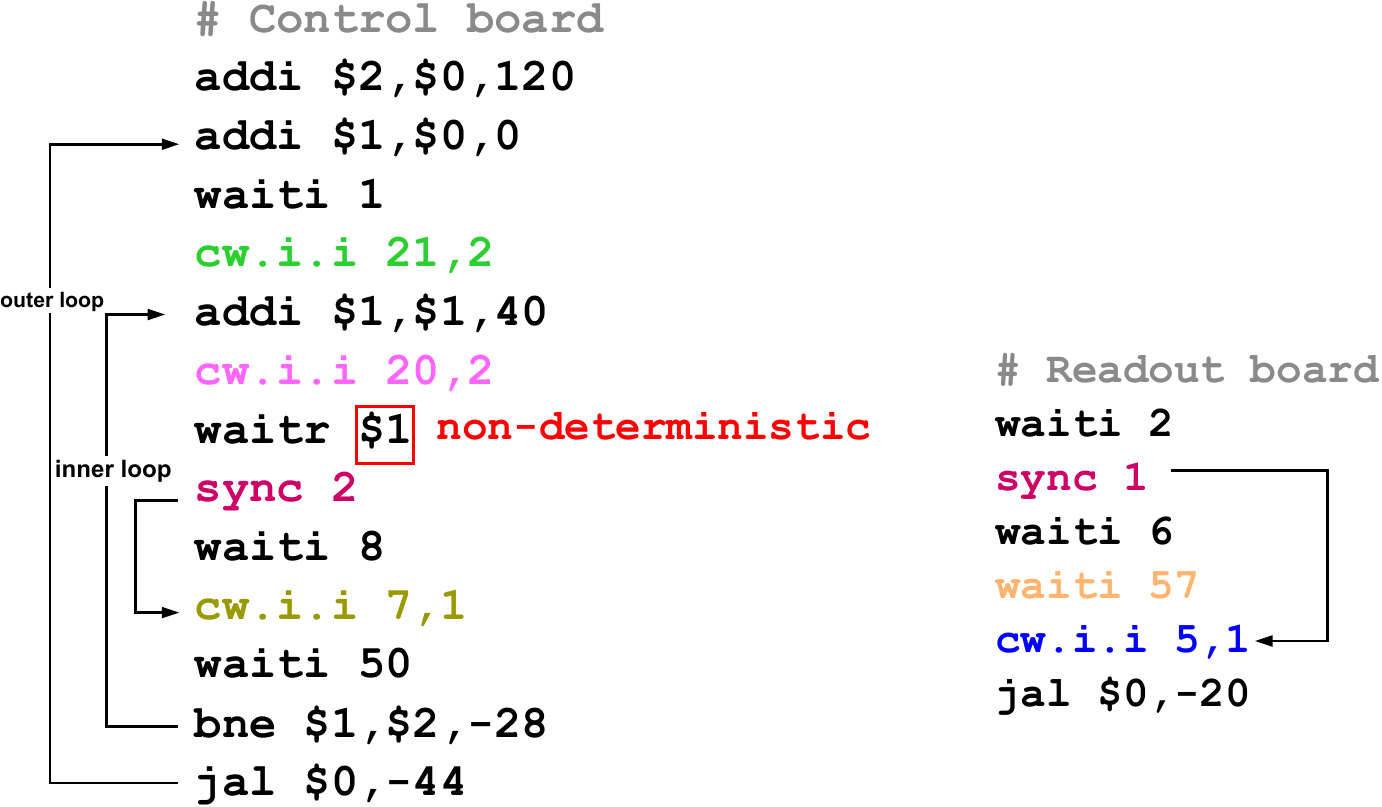}
    \caption{HISQ instructions running on the control and readout boards. Note that since the two boards have different triggering delays, we added a 57-cycle delay (the \code{waiti 57}) before the synchronous operation of readout board to mitigate this difference.}
    \label{fig:eva_insn}
\end{figure}

We designed two programs (Figure~\ref{fig:eva_insn}) that respectively run on a control board and a readout board to verify the feasibility of \sync{}.
Both boards repeatedly execute sync instructions.
Except for the sync instruction, the readout board contains only deterministic tasks, while the control board contains a non-deterministic task -- the \iname{waitr \$1} instruction.
The varied timing of control board makes its progress unpredictable to the readout board.
As such, we emulated the as-needed synchronization scenario in DQCA.

\begin{figure}[bt]
    \centering
    \includegraphics[width=\linewidth]{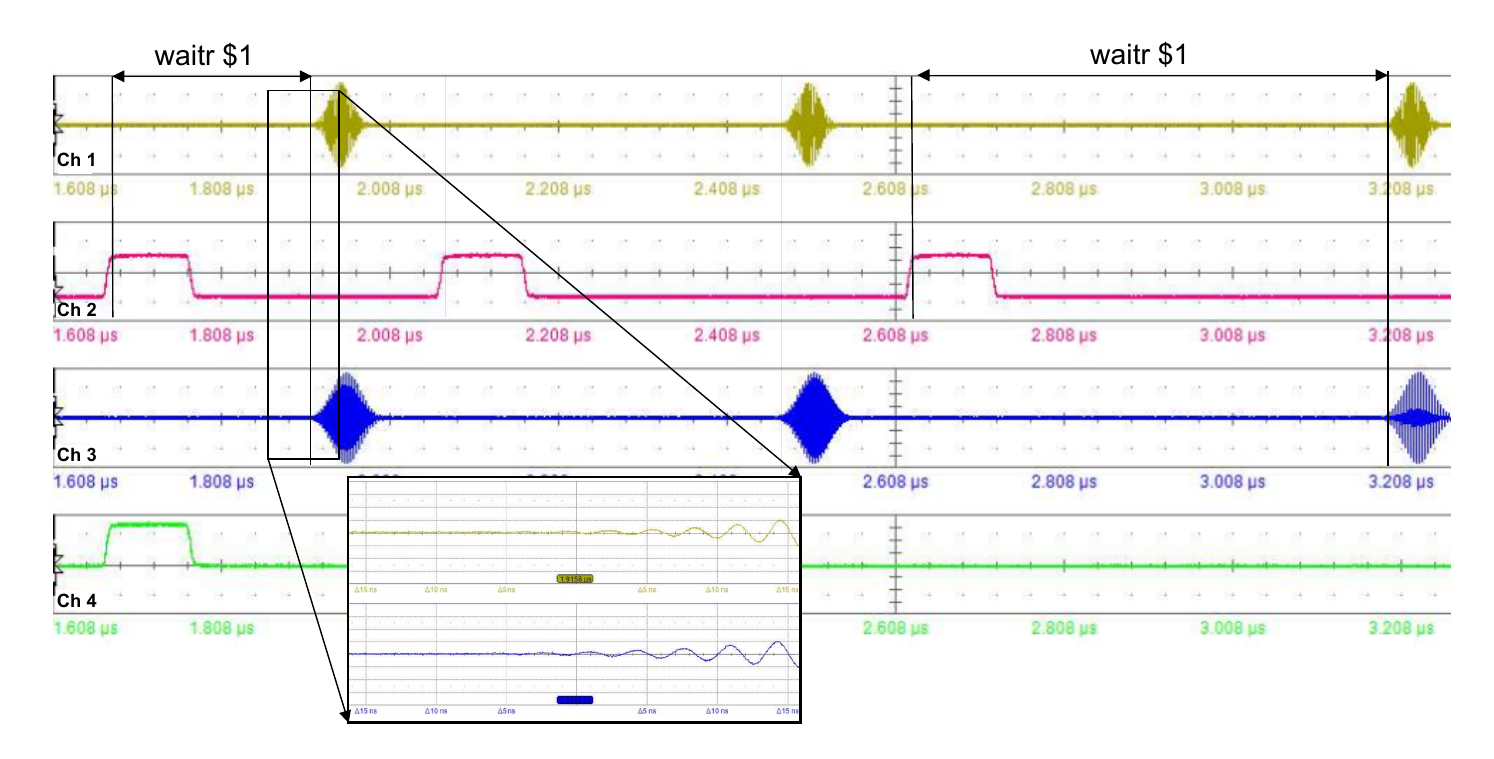}
    \caption{Waveforms illustrating synchronization between nearby control board and readout board.}
    \label{fig:eva_waveform}
\end{figure}

The result of the instruction execution is shown in the waveform diagram (Figure ~\ref{fig:eva_waveform}).
Channels 1 and 2 of the oscilloscope reveal that the start time of the control board’s sync increases by 120 ns in each inner loop iteration, reflecting the increment of the register value \iname{\$1}.
The instructions requiring synchronization are highlighted in yellow and blue (Figure~\ref{fig:eva_insn}) on both the control and readout boards, corresponding to the yellow and blue pulses in Figure~\ref{fig:eva_waveform}. As shown, regardless of changes in \iname{\$1}, the yellow and blue instructions are always executed synchronously at the cycle level.

\subsection{Simulation-Level Verification}
\label{ssec:sim-eval}

\subsubsection{Simulation Platform}

To enable efficient evaluation, we also developed a simulator, CACTUS-Light, based on the open-source QCA simulator CACTUS~\cite{zhang2018QuMAsimQuantumArchitecture,cactus}, but with the microarchitecture under investigation modeled at transaction level.
CACTUS-Light adds support for the synchronization module as proposed in Section~\ref{sec:sync-protocol}.
It has been verified at two levels.
The logical correctness is verified using multiple small-scale benchmarks whose execution produces expected quantum state or measurement results.
The timing information in the simulation result is verified against the FPGA implementation using Timing Event Logging Format (TELF)~\cite{zhang2018QuMAsimQuantumArchitecture} data.
In our evaluation, we set \SI{20}{\ns} (\SI{40}{\ns}) for single (two)-qubit gates, and \SI{300}{\ns} for measurements.

\begin{figure}
    \centering
    \includegraphics[width=\linewidth]{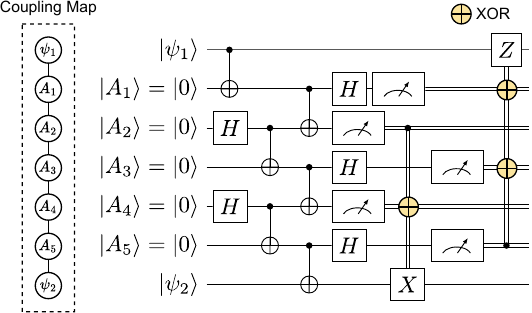}
    \caption{Long-range CNOT gate between $|\psi_1\rangle$ and $|\psi_2\rangle$ based on gate-teleportation~\cite{baumerEfficientLongRangeEntanglement2024}. Although SWAP gates can enable a CNOT between $|\psi_1\rangle$ and $|\psi_2\rangle$, the circuit depth increases linearly with qubit count. In contrast, this scheme maintains constant circuit depth as the number of qubits grows.}
    \label{fig:long-range-cnot}
\end{figure}

\subsubsection{Benchmarks}

We constructed two types of benchmarks.

\begin{enumerate}
    \item \tb{Near-term dynamic circuits.}
    Qubit connectivity constraints in quantum devices have spurred extensive research into qubit mapping and routing challenges~\cite{liTacklingQubitMapping2019,tanOptimalLayoutSynthesis2020}.
    By supporting arbitrary feedback operations, these constraints can, in principle, be overcome using non-unitary dynamic circuits~\cite{baumerEfficientLongRangeEntanglement2024}.
    Figure~\ref{fig:long-range-cnot} illustrates a circuit diagram for implementing a long-range CNOT between two distant qubits via dynamic circuits.
    This approach trades spatial resources for temporal efficiency, utilizing additional ancilla qubits to eliminate cumbersome SWAP gates and achieve reduced circuit depth.
    Based on this approach, we have converted several static circuits from QASMBench~\cite{liQASMBenchLowLevelQuantum2023} to dynamic circuits
    by randomly substituting CNOTs between non-adjacent qubits with long-range CNOTs.
    Subsequently, we convert these OpenQASM~\cite{crossOpenQASM3Broader2022} programs into HISQ programs to serve as benchmarks.

    \item \tb{Logical $T$ gate-based QEC circuits.}
    With the long-term goal of achieving FTQC, various QEC protocols have been developed, with the surface code recognized as a leading candidate~\cite{fowlerSurfaceCodesPractical2012,fowlerLowOverheadQuantum2019}.
    Logical $T$ gates are both resource-intensive and frequent in this protocol~\cite{fowlerSurfaceCodesPractical2012,yinSurfDeformerMitigatingDynamic2024}. Implementing a logical $T$ gate involves a logical feedback operation (Figure~\ref{fig:logical_t}), making it an ideal case to evaluate \name{}’s performance in QEC experiments.
    We construct and validate logical $T$ gates using lattice surgery~\cite{horsmanSurfaceCodeQuantum2012a} with Stim~\cite{gidneyStimFastStabilizer2021}, then convert these circuits into HISQ programs.
    As our focus is synchronization efficiency, we do not implement error decoding, but model its latency by inserting \iname{wait} instructions based on existing hardware decoder data~\cite{barberRealtimeScalableFast2025a}, assuming each \router{} has a dedicated decoder.
    Given the high overhead of magic state distillation, we assume pre-prepared magic states and only simulate the logical feedback portion of the $T$ gate.
\end{enumerate}

\subsubsection{Baseline}

As the experiment baseline, we implemented the lock-step synchronization scheme as proposed in ~\cite{zettles262DesignConsiderations2022,ibm_central_controller}.
The control microarchitecture of each controller is identical to that of \name{}.
A central controller orchestrates the execution of all other controllers with a star topology.
This approach limits the flexibility of feedback operation execution, particularly under conditions with numerous concurrent feedback operations.
In contrast, our proposed scheme enables asynchronous execution of concurrent feedback operations, aligning them only when required.
In simulation, we assume unlimited connectivity for the baseline and treat the communication latency of a feedback operation as constant, regardless of the number of qubits.
In practice, however, this assumption is unrealistic, and thus our simulation results actually overestimate the performance of the baseline.

\begin{figure}
    \centering
    \includegraphics[width=\linewidth]{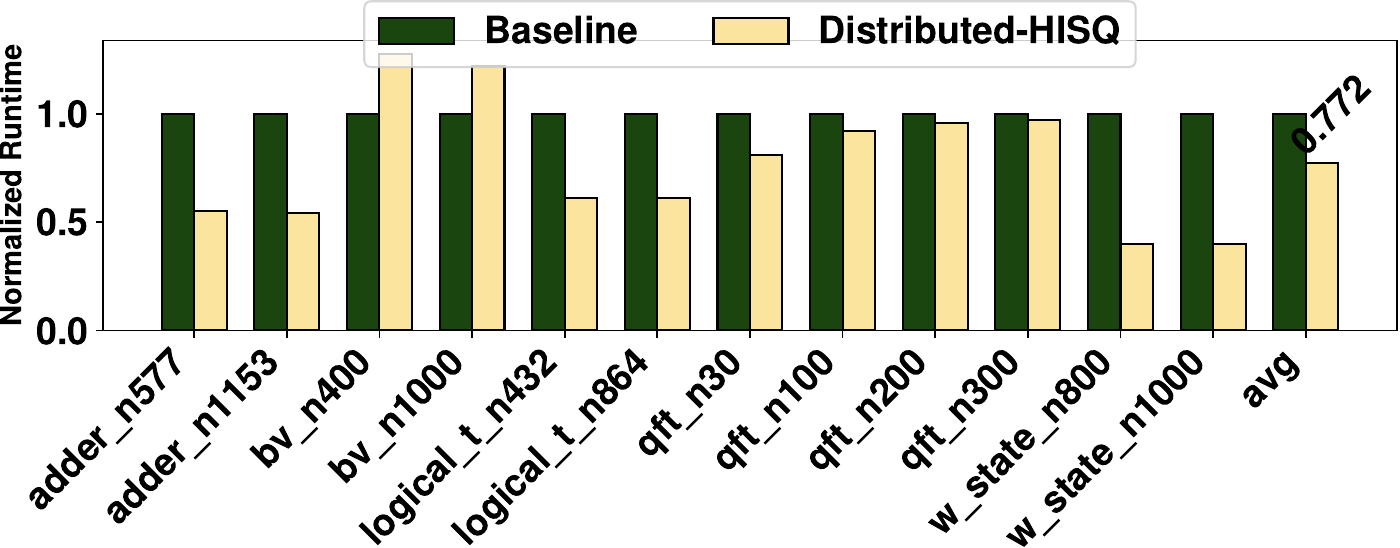}
    \caption{Runtime comparison with baseline. These benchmarks are dynamic circuits obtained by compiling static circuits with long-range CNOT gates~\cite{baumerEfficientLongRangeEntanglement2024}.}
    \label{fig:comp_baseline}
\end{figure}

\subsubsection{Results Analysis}
\label{sssec:res_analysis}

Figure~\ref{fig:comp_baseline} shows the normalized end-to-end runtime of a HISQ program, demonstrating that \name{} reduces execution time by an average of 22.8\%.
The reason is that there is no simultaneous feedback in this benchmark, making the advantage of \name{} not significant.
On the other hand, the communication latency of \name{} grows as the system scale up, but we assume constant communication latency in baseline.
As such, the performance of \name{} is worse than baseline for ``bv'' benchmark.

\subsubsection{Impact on Fidelity}

\begin{figure}
    \centering
    \includegraphics[width=\linewidth]{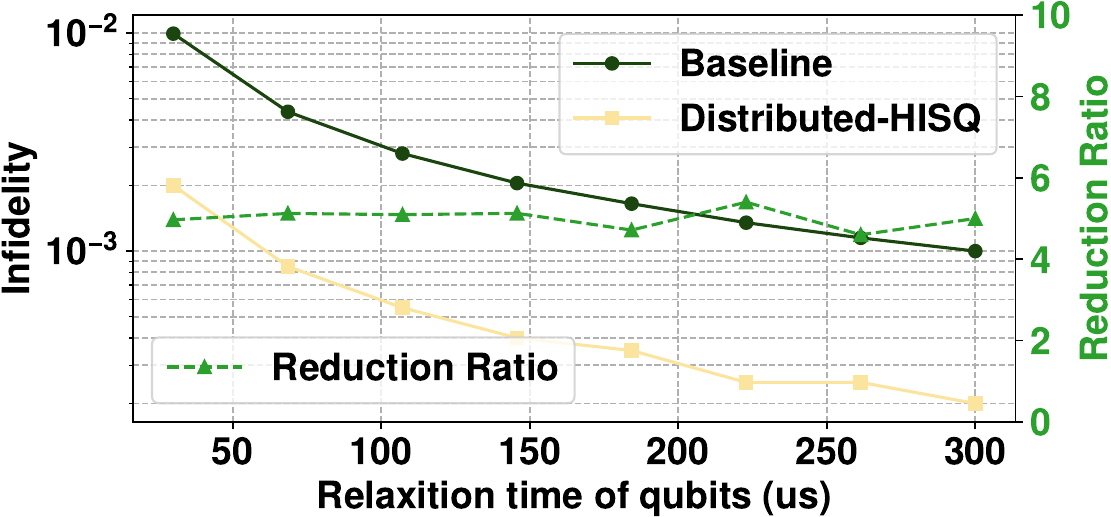}
    \caption{Fidelity comparison between \name{} and baseline.}
    \label{fig:fidelity}
\end{figure}

As an example to showcase the effect of our reduced latency on fidelity, we consider the long-range CNOT circuit shown in Figure~\ref{fig:long-range-cnot}.
We compare the infidelity between \name{} and baseline with the T1/T2 time ranging from \SIrange{30}{300}{\us} (Figure~\ref{fig:fidelity}).
It can be observed that \name{} constantly reduces infidelity around 5$\times$.
This effect is also due to the ability to enable simultaneous feedback. In the baseline scheme, all controllers are forced to follow the same program flow, requiring the second set of measurements to occur after the conditional $X$ gate. In contrast, \name{} allows these measurements to be performed immediately after the $H$ gates, thereby reducing execution time.

\section{Discussion}
\label{sec:discussion}
\subsection{Adaptability of HISQ}

While our current single-core DQCtrl implementation --- where one HISQ core controls all 28 ports on an FPGA --- is adequate for many tasks, it faces a potential instruction issue rate bottleneck~\cite{fuEQASMExecutableQuantum2019} in time-critical operations. 
To overcome this scalability problem, our architecture allows for a multi-core configuration on a single FPGA. By partitioning the control ports among multiple HISQ cores, we eliminate the issuance bottleneck, ensuring robust performance for computationally demanding experiments.

\subsection{Scalability of \name{}}

The scalability of a quantum control architecture is determined by various factors, including instruction issue rate, memory consumption, and synchronization latency, and is ultimately limited by its most critical bottleneck. While this work does not propose a universally scalable quantum control architecture, it targets two key challenges: achieving efficient synchronization and enabling a lightweight hardware implementation for the digital control logic. 
\name{} utilizes a queue-based event timing mechanism that compiles a single quantum program into independent instruction streams for multiple controllers. These streams execute in parallel and synchronize only on demand, significantly reducing the instruction load on each controller. This partitioned execution model enhances the efficiency of individual controllers, thereby improving the overall scalability of the quantum control architecture.

\section{Conclusion}
\label{sec:conc}

We have presented \name{}, a distributed quantum control architecture addressing key challenges in scalability and adaptability for evolving quantum hardware. 
Its core contributions are twofold: a hardware-implementable yet expressive abstraction layer, and an efficient synchronization scheme that provides near-zero overhead and enhances compiler flexibility. 
We implemented and verified the architecture on a realistic superconducting qubit control system. Furthermore, simulation results demonstrate that \name{} reduces execution overhead and improves result fidelity. By effectively decoupling instruction execution while maintaining precise control, \name{} provides a viable pathway toward fully scalable quantum control architectures.

\begin{acks}
    We thank the anonymous reviewers for their insightful feedback.
    We thank Dr. Lianchen Han for his support in debugging the firmware of DQCtrl.  
    This work was supported in part by National Natural Science Foundation of China (Grant No. 62025404, 62222411), National Key Research and Development Program of China (Grant No. 2023YFB4404400).
\end{acks}

\bibliographystyle{ACM-Reference-Format}
\bibliography{refs}

\end{document}